\documentclass[11pt,twoside]{article}
\usepackage{asp2004}
\usepackage{psfig}
\usepackage{epsf}
\usepackage{graphics}
\usepackage{lscape}
\begin{document}
\title{Multiwavelength WHAM Observations of Extra-planar Warm Ionized Gas in the Galaxy}
\author{G.~J.~Madsen}
\affil{Department of Astronomy, University of Wisconsin - Madison, 475 N. Charter Street, Madison, WI 53706, USA}

\begin{abstract}
We report on observations of several optical emission lines toward a variety of newly revealed faint, large-scale H$\alpha$-emitting structures in the warm ionized medium (WIM) of the Galaxy. The lines include [\ion{N}{II}]$~\lambda6583$, [\ion{N}{II}]$~\lambda5755$, [\ion{S}{II}]$~\lambda6716$, [\ion{O}{III}]$~\lambda5007$, and \ion{He}{I}$~\lambda5876$\ and were obtained with the Wisconsin H-Alpha Mapper (WHAM) toward sightlines that probe both large filamentary features and the more diffuse WIM, from the outer Perseus spiral arm to the inner Galaxy.   
These emission lines allow us to explore the temperature and ionization conditions within the emitting gas and their variations between the different emission regions.
We compare the relative intensities of the emission lines toward these faint \ion{H}{II}\ structures to  brighter, classical O star \ion{H}{II}\ regions.
We find that the line ratios of [\ion{N}{II}]/H$\alpha$\ and [\ion{S}{II}]/H$\alpha$\ are generally high, while [\ion{O}{III}]/H$\alpha$\ and \ion{He}{I}/H$\alpha$\ are generally low in the diffuse WIM and in the faint filamentary structures. This suggests that the gas is warmer, in a lower ionization state, and ionized by a softer spectrum than the gas in classical \ion{H}{II}\ regions surrounding O stars, the presumed ionization source for the WIM. In addition, we find differences in the physical conditions between the large filamentary structures and the more diffuse WIM, suggesting that the filaments are regions of higher density, not geometrical projections of folds in large sheet-like or shell-like structures. 

\end{abstract}
\thispagestyle{plain}

\section{Introduction}
Within the past 20 years, the warm ionized medium (WIM) has emerged as an important component of the interstellar medium (ISM) of our Galaxy \citep{Reynolds90-IAU}. Characterized by its low density (0.1 cm$^{-3}$), large scale height (1 kpc), and warm temperature (10$^4$~K), the diffuse WIM pervades the Galaxy and significantly impacts our understanding of heating and ionization processes in the ISM. Despite its importance, the nature and origin of the WIM remains poorly understood. The WIM is thought to be photoionized by O and early B stars that are principally located near the Galactic plane. However, this requires that a significant fraction of Lyman continuum radiation from these stars is able to propagate through the ubiquitous neutral \ion{H}{I}\ disk and up into the halo. It is unlikely that most of the WIM is ionized through shocks, because this requires that supernovae convert almost all of their mechanical energy into Lyman continuum radiation, and is inconsistent with the narrow line widths ($\sim$ 20 ${\rm km\thinspace s^{-1}}$) of emission lines in the WIM.  Photoionization models of the WIM employ low ionization parameters to reproduce the observations, but still have difficulty reproducing all of the emission line ratios \citep[e.g.,][see also Wood; Elwert, this volume]{CR01}. In addition, recent observations suggest that a significant heating source, beyond photoionization, may become dominant in low density regions of the WIM where $n_e \la$ 0.1 cm$^{-3}$ \citep{RHT99}.

One of the primary sources of information about the WIM comes from studies of faint optical emission lines. The Wisconsin H-Alpha Mapper (WHAM) has observed the entire northern sky ($\delta >$ -30$\deg$) in the Balmer-$\alpha$ line, and has produced, for the first time, velocity-resolved maps of \ion{H}{II}\ with a spectral and spatial resolution comparable to all-sky \ion{H}{I}\ maps \citep{Haffner+03}. The WHAM sky survey has revealed the presence of bright, classical \ion{H}{II}\ regions along with fainter large-scale ionized filaments, loops, and arcs, superposed on a diffuse background.  The physical conditions of the WIM can be studied through the use of nebular emission line diagnostics that are commonly used for much brighter sources (e.g. \ion{H}{II}\ regions and planetary nebulae). 
The WHAM survey provides an excellent basis for a similar study of the WIM.  Comparing the relative strengths of various emission lines allows us to explore the range of physical conditions present among the various features in the WIM, search for patterns in these conditions, and relate these conditions to those associated with classical \ion{H}{II}\ regions. Here, we discuss such a study toward several interesting features in the WIM.

\section{Observations}

All of our observations were obtained with the WHAM instrument, which is a dual-etalon Fabry-Perot spectrometer that produces an optical spectrum with a resolution of 12 ${\rm km\thinspace s^{-1}}$\ over a 200 ${\rm km\thinspace s^{-1}}$\ spectral window, averaged over its 1$\deg$\ field of view. WHAM was specifically designed to study faint, diffuse emission with high sensitivity, and can observe lines as faint as $\sim$ 0.01 R\footnotemark\footnotetext{1 R = $10^6$/4$\pi$ photons s$^{-1}$ cm$^{-2}$ sr$^{-1}$}.  We have observed of several regions of the Galaxy in the lines of H$\alpha$, H$\beta$, [\ion{N}{II}]$~\lambda6583$, and [\ion{S}{II}]$~\lambda6716$, as well as the much fainter lines of [\ion{N}{II}]$~\lambda5755$, [\ion{O}{III}]$~\lambda5007$, and \ion{He}{I}$~\lambda5876$. These observations provide an order of magnitude increase in the number of detections of \ion{He}{I}\ and [\ion{O}{III}]\ in the WIM. In \S\S3-6 below, we discuss sets of observations toward a few distinguished features in the WIM, followed by a  summary of our results in \S7. A more complete discussion of these observations can be found in \citet{MadsenPhD}.

\section{Orion-Eridanus Bubble}

\begin{figure}[!ht]
\plotone{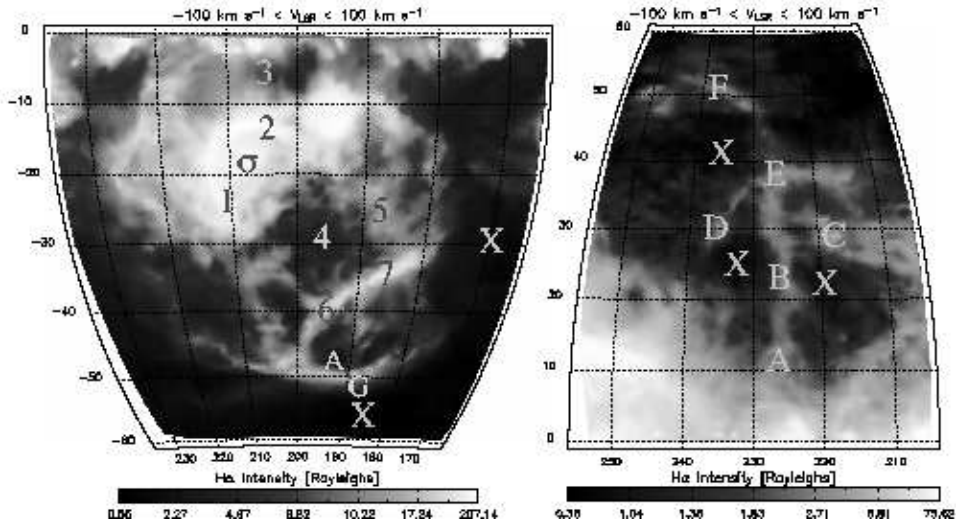}
\caption{H$\alpha$\ maps of the Orion-Eridanus Bubble ({\it{left}}) and the Northern Filament ({\it{right}}).}
\vspace*{-0.2in}
\end{figure}

One of the largest networks of interconnected H$\alpha$-emitting structures in the WHAM H$\alpha$\ sky survey is in the constellations of Orion and Eridanus, shown on the left in Figure 1.  \citet{RO79} combined kinematic and spatial emission-line data toward this region, and found that the filaments and loops are all part of an asymmetrically expanding shell of neutral and ionized gas, with a diameter of $\approx 280$ pc. They suggested that Lyman continuum photons travel largely unimpeded through this hot ($T \sim 10^6$~K) cavity, ionizing its walls, and is consistent with the detection of diffuse X-ray emission interior to the bubble walls. Among the most luminous, hot stars within the bubble, $\delta$ Ori, an O9.5I star, has no \ion{H}{II}\ region around it, implying that most of its ionizing radiation can travel unimpeded through the cavity and ionize the bubble walls.

\begin{figure}[!ht]
\plotone{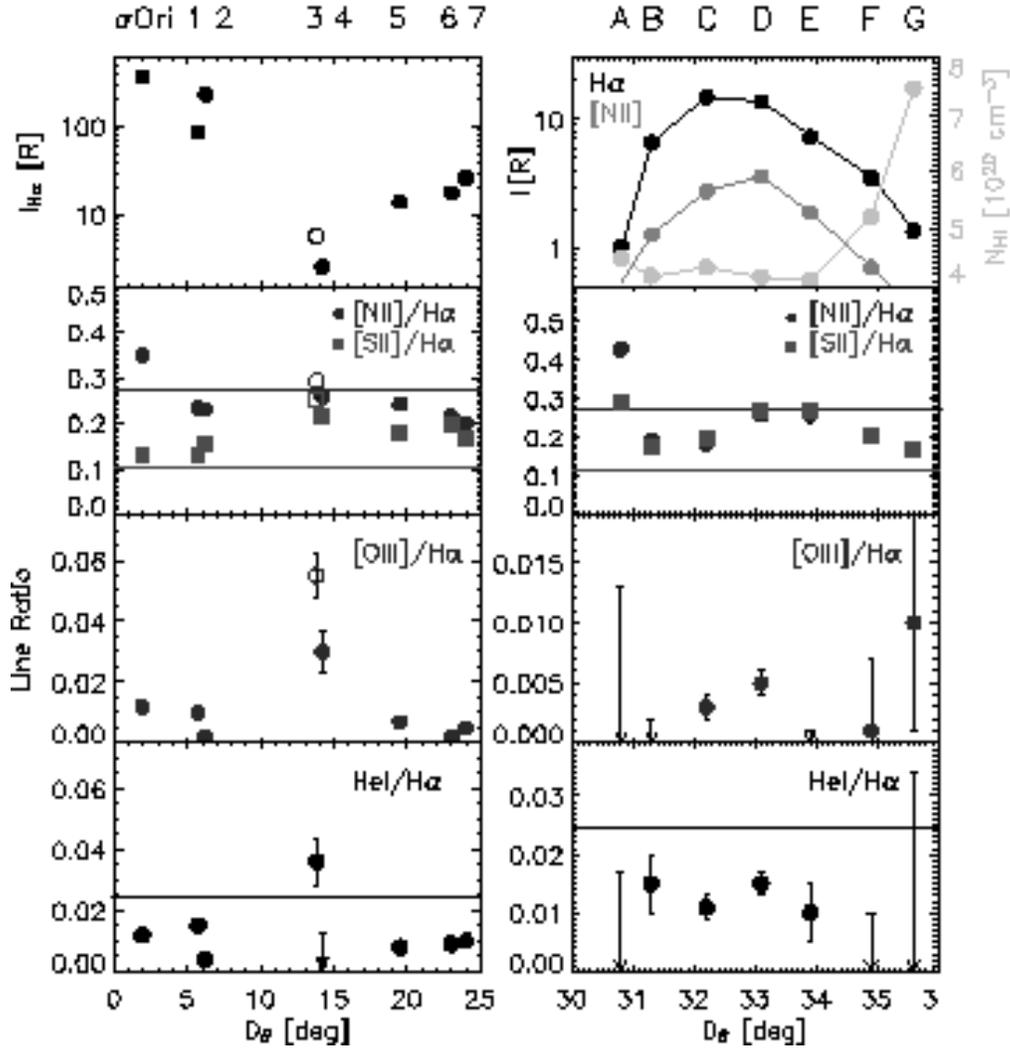}
\caption{Comparison of emission line strengths toward the Orion-Eridanus Bubble. The line strengths and ratios are shown as a function of angular distance from the center of the Orion OB1 association. The left panel includes the $\sigma$ Ori \ion{H}{II}\ region, as well as directions in and near the Bubble. The right panel includes directions, spaced 1$\deg$\ apart, that cut across the southern edge of the Bubble. 
The column density of \ion{H}{I}, from \citet{HIAtlas}, is also shown in light grey.
The average line ratios for several classical \ion{H}{II}\ regions are shown as solid horizontal lines.}
\end{figure}

A summary of our multiwavelength emission line observations toward this bubble is shown in Figure 2. The letter and numbers at the top of the plot refer to the labels shown in Figure 1. The line intensities and ratios are shown as a function of angular distance, $D_\theta$, from the approximate center of Orion OB1 assocation.  The variations in [\ion{N}{II}]/H$\alpha$\ can be attributed to variations in temperature of the emitting gas, since N$^+$/H$^+$ $\approx$ 1 \citep{HRT99}. We see that [\ion{N}{II}]/H$\alpha$\ is $\approx 0.25$, which is near the average value for several classical \ion{H}{II}\ regions.  We see no strong trends with $D_\theta$. 
The line ratio [\ion{S}{II}]/[\ion{N}{II}]\ is nearly independent of temperature, and variations in [\ion{S}{II}]/[\ion{N}{II}]\ can be attributed to variations in S$^+$/S \citep{HRT99}. 
Compared to the average value of [\ion{O}{III}]/H$\alpha$\ $\approx 0.09$ for several classical \ion{H}{II}\ regions, [\ion{O}{III}]/H$\alpha$\ in and around this bubble is very low, indicating a relatively low ionization state of the gas. 
The low \ion{He}{I}/H$\alpha$\ measurements suggest that He$^+$/He $\la$ 0.3, which is consistent with an ionizing spectrum from a continuum source with $T_* \la$ 35,000 K, equivalent to a O8.5I or O9.5V star or cooler.
Across the southern edge of the shell, directions A-G, we find that the [\ion{N}{II}]\ and [\ion{S}{II}]\ emission is strongest behind the H$\alpha$\ filaments, further from the ionizing source, but is interior to the location of the brightest \ion{H}{I}\ emission.

\section{Perseus Superbubble}

\begin{figure}[!ht]
\plotone{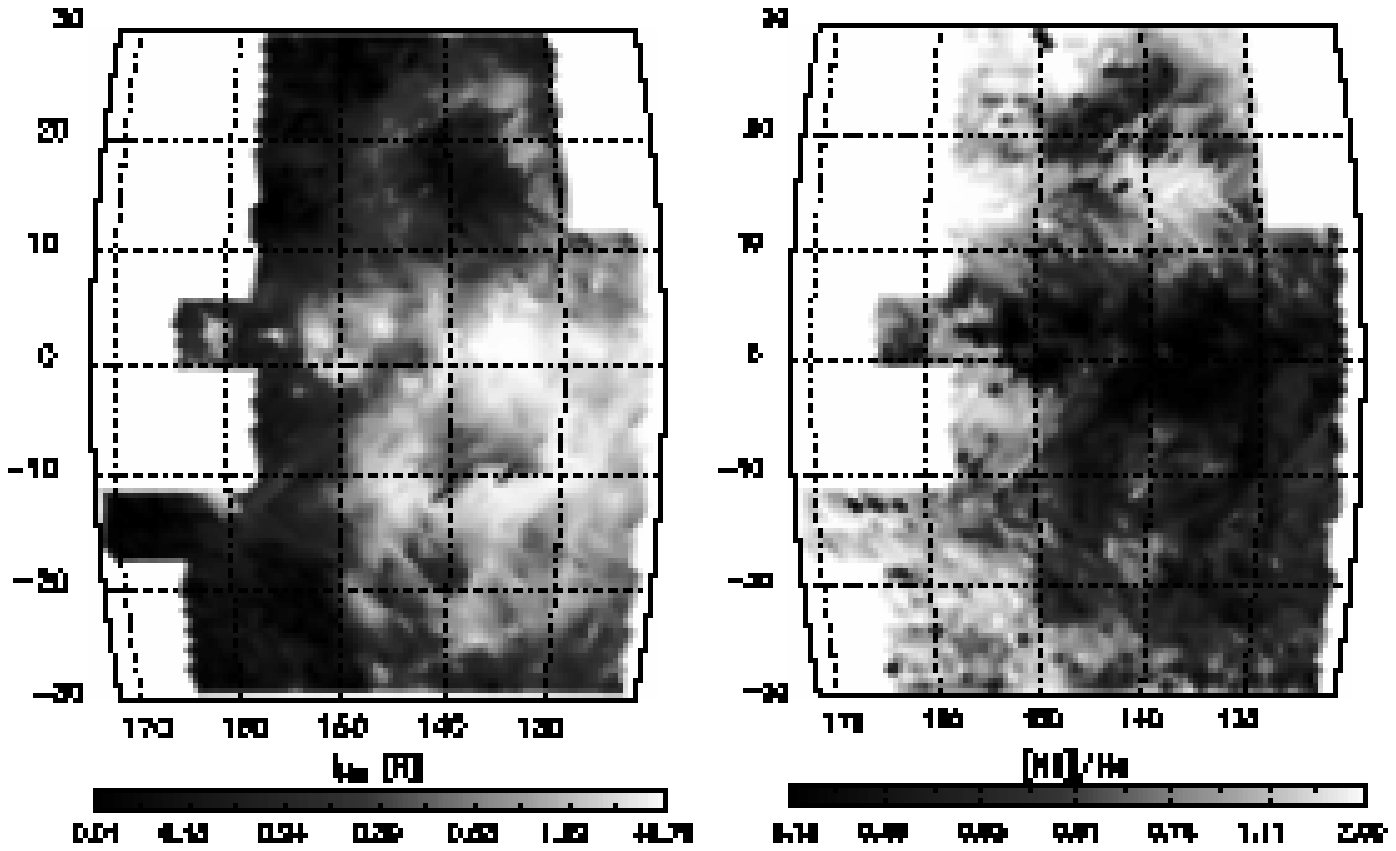}
\caption{Maps of H$\alpha$\ ({\it{left}}) and [\ion{N}{II}]/H$\alpha$\ ({\it{right}}) toward the Perseus Superbubble. Note the anti-correlation between H$\alpha$\ intensity and [\ion{N}{II}]/H$\alpha$\ line ratio.}
\end{figure}

Another, much larger, bubble-like structure is shown in Figure 3. The map on the left is of H$\alpha$\ emission at velocities $-75\ {\rm km\thinspace s^{-1}} < v_{LSR} < -45\ {\rm km\thinspace s^{-1}}$, corresponding to emission from the Perseus spiral arm at a distance of $\approx$ 2 kpc. A bipolar loop structure extends up to 30$\deg$, or $\approx$ 1 kpc, above and below the Galactic plane.
The W4 star-forming region is near the center of this structure at $(\ell,b)$\ $\approx$ (135$\deg$, 0$\deg$), and several studies have suggested that a large `chimney' has been carved out there, allowing radiation and hot gas to move out into the Galatic halo \citep[e.g.,][]{NTD96, RSH01}. 
We observed [\ion{N}{II}]\ and [\ion{S}{II}]\ emission throughout this entire region. The morphology of the [\ion{N}{II}]\ and [\ion{S}{II}]\ maps are very similar to the H$\alpha$\ map. However, the ratio of [\ion{N}{II}]/H$\alpha$, shown on the right in Figure 3, shows several interesting trends. 
We see a detailed anti-correlation between H$\alpha$\ intensity and [\ion{N}{II}]/H$\alpha$\ line ratio that holds on large and small spatial scales, with regions of low emission measure having an elevated [\ion{N}{II}]/H$\alpha$\ ratio. 
This suggests that regions of enhanced emission measure are cooler, and are likely regions of higher density.
This relationship has been observed elsewhere when comparing bright \ion{H}{II}\ regions with the diffuse WIM \citep{HRT99}. Here we see that this trend holds even among individual features {\it{within}} the WIM. 

\section{Northern Filament}

The map on the right in Figure 1 shows a remarkable, $\sim 2\deg$-wide, H$\alpha$\ filament that rises vertically 60$\deg$\ from the Galactic plane near longitude $l=225\deg$, which we refer to as the `Northern Filament'. 
\citet{HRT98} examined this region in the H$\alpha$\ survey and found that the filament is likely associated with the \ion{H}{II}\ region S292 near $(\ell,b)$\ $\approx$ (225$\deg$, 0$\deg$).
However, they did not find an acceptable explanation for its origin, and point out that the relatively constant H$\alpha$\ surface brightness along the filament suggests that it is ionized by ambient Lyman continuum radiation escaping the Galactic midplane, rather than from a jet-like phenomenon.
We have observed several emission lines toward the labeled directions A-F overlayed on the map in Figure 1.
We find that along the length of the filament, [\ion{N}{II}]/H$\alpha$\ remains constant, $\approx$ 0.4 - 0.5, suggesting that the temperature is $T \approx$\ 7400 K, similar to what is found in other regions of the WIM with comparable emission measure.  This implies that the appearance of this filament in the H$\alpha$\ sky is due to a density enhancement, the same conclusion reached regarding the filaments in the Perseus arm.   
The constancy of H$\alpha$, [\ion{N}{II}]/H$\alpha$, and [\ion{S}{II}]/H$\alpha$\ also implies that the filament is not likely to have been impulsively ionized, since these ions have significantly different recombination timescales.
In [\ion{O}{III}], a very strong trend is seen, with [\ion{O}{III}]/H$\alpha$\ $\approx 0.06$ for S292 and for direction A, and falls to $\la 0.01$ for the two highest directions. This is opposite to what is generally seen in external galaxies \citep[e.g.,][]{CR01}, as well as in the inner Galaxy (\S6 below), where [\ion{O}{III}]/H$\alpha$\ is observed to increase with increasing distance from the plane.

\section{Inner Galaxy}

A suprising result from the H$\alpha$\ sky survey was the detection of H$\alpha$\ emission toward a region of the inner Galaxy at velocities that exceeded $v_{\rm{LSR}} \ga$ +100 ${\rm km\thinspace s^{-1}}$, the positive velocity limit of the survey. 
We have reobserved this region of the Galaxy and have indentified a strong concentration of high velocity emission ($+100~{\rm km\thinspace s^{-1}} \la v_{\rm{LSR}} \la +150~{\rm km\thinspace s^{-1}}$) toward a $\approx 5\deg \times 5\deg$ area centered near $(\ell,b)$ $= (27\deg,-3\deg)$, known as the Scutum Cloud. 
The ratio of H$\alpha$/H$\beta$\ as a function of velocity suggests that $A(V) \approx$ 3 out to the tangent point velocity, and suggests that we are detecting optical emission a few degrees away from the Galactic plane at a distance of more than 6 kpc from the Sun. Figure 4 shows spectra of H$\alpha$, H$\beta$, [\ion{N}{II}], [\ion{S}{II}], and [\ion{O}{III}]\ toward this low extinction window, with the data averaged over four latitude bins.
We find that [\ion{N}{II}]/H$\alpha$\ and [\ion{S}{II}]/H$\alpha$\ both increase with increasing distance from the plane, suggesting that the temperature increases by $\approx$ 1000 K, at a given radial velocity/heliocentric distance. 
At a given latitude, [\ion{S}{II}]/[\ion{N}{II}]\ tends to decrease with increasing velocity, suggesting that S$^{+}$/S is lower in the inner Galaxy, where S$^{++}$/S is presumably higher. This trend of increased ionization state closer to the Galactic center is consistent with the [\ion{O}{III}]/H$\beta$\ data, where the data suggest that O$^{++}$/H$^+$ increases dramatically, and is consistent with more star formation in the inner Galaxy.

\begin{figure}[!ht]
\plotone{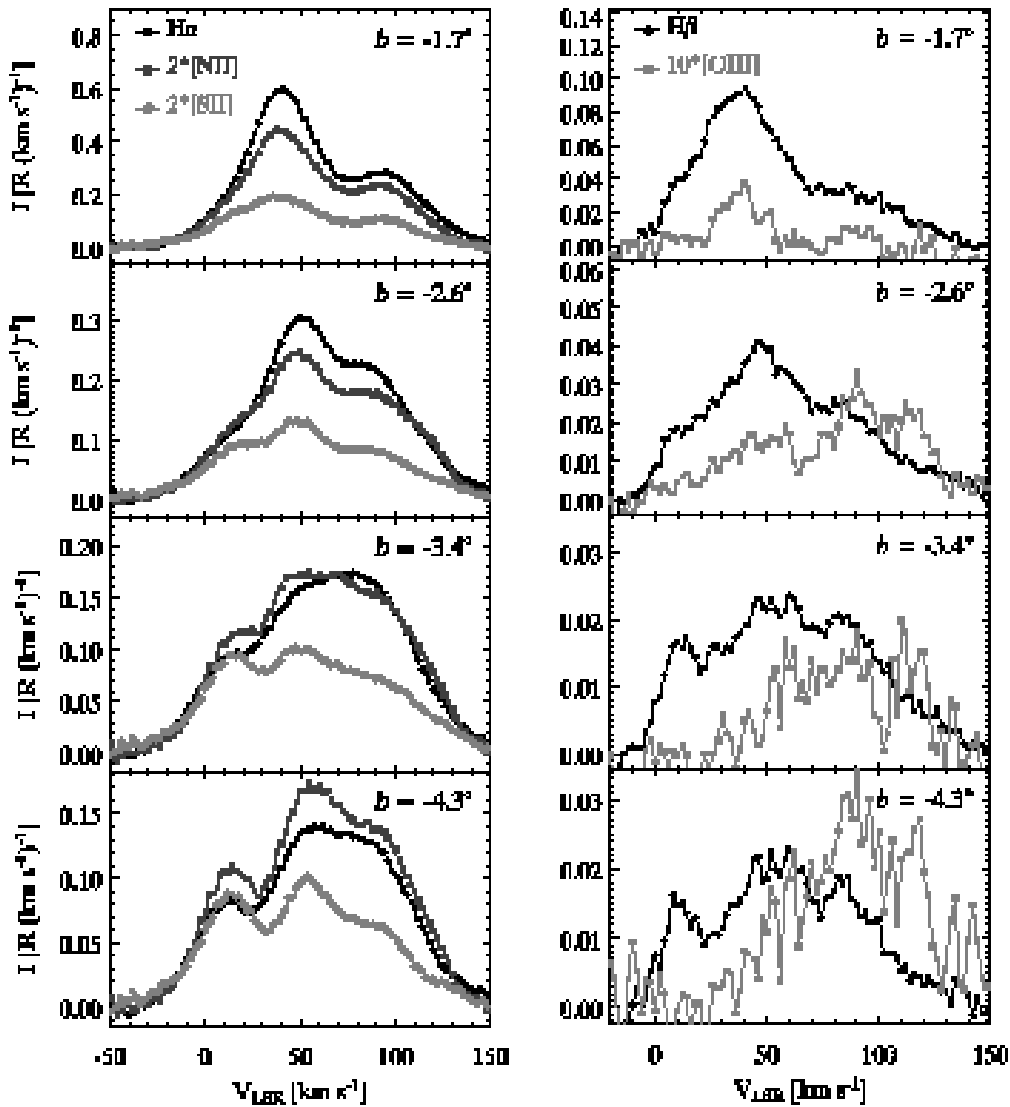}
\caption{Multiwavelength spectra toward a low extinction window in the inner Galaxy.}
\vspace*{-0.2in}
\end{figure}

\section{Summary}

\begin{figure}[!ht]
\plotone{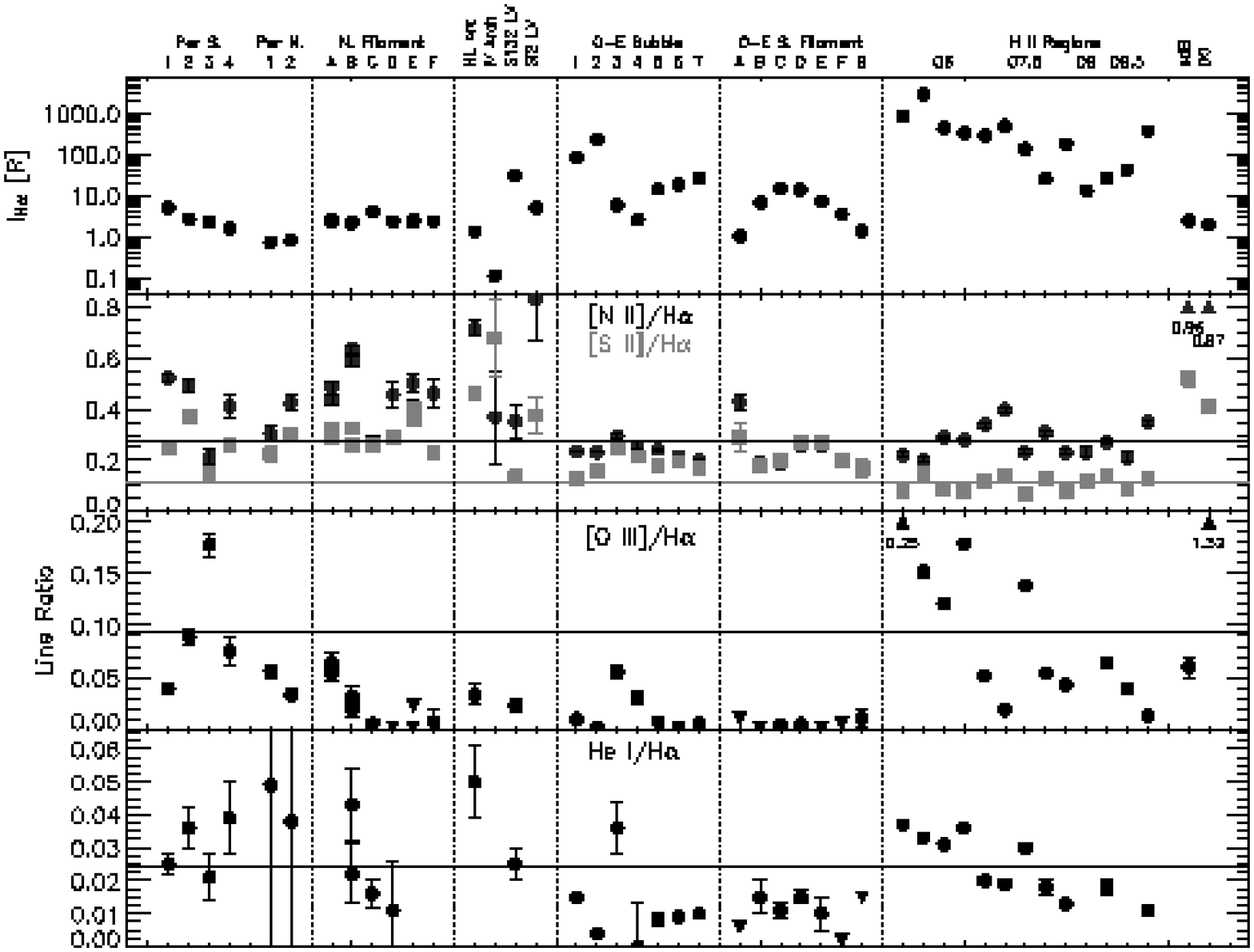}
\caption{Summmary of all emission line pointed observations. Data for different kinds of features are separated by vertical dashed lines. }
\vspace*{-0.1in}
\end{figure}

Figure 5 shows all of our pointed observations of multiple emission lines toward several interesting features in the WIM. These include multiple velocity components toward the northern and southern loop of the Perseus superbubble, the Northern Filament, a high latitude arc, part of an \ion{H}{I}\ feature known as the  IV Arch, the Orion-Eridanus bubble, several classical \ion{H}{II}\ regions, and two hot evolved stellar cores. Large-scale maps in H$\alpha$, [\ion{N}{II}], and [\ion{S}{II}]\ toward the Orion and Perseus bubbles, over different velocity intervals, have also been analyzed. From these data, we draw the following conclusions:

1) The temperature of warm ionized gas is higher in regions of lower emission measure. The higher temperatures in the WIM are confirmed by the [\ion{N}{II}]$~\lambda5755$/[\ion{N}{II}]$~\lambda6583$ intensity ratios.

2) Filamentary structures have physical conditions that differ from those in the fainter, more diffuse background. This implies that they are likely regions of enhanced density, not directions of increased pathlength through the diffuse background along folds or edge projections of thin shells or sheets. 

3) The ionization state in the WIM is lower than in \ion{H}{II}\ regions.
We find that the fraction of O$^{++}$/O and He$^+$/He in the WIM is low compared to  \ion{H}{II}\ regions, implying a lower ionization state and softer ionizing radiation field. Low density gas closer to the Galactic plane may be more highly ionized than gas at larger distances, with this trend reversing in the inner Galaxy.

4) Physical conditions within large ionized bubbles do not change significantly with distance from the ionizing source.
Compared to the Perseus superbubble, the smaller, brighter Orion-Eridanus bubble has low [\ion{N}{II}]/H$\alpha$\ ratios, similar to those in classical \ion{H}{II}\ regions.
This implies that bubble size, gas density within the ionized shell, or the flux and spectrum of the radiation escaping the O star cluster may  be important in setting the conditions within the ionized gas. 

\acknowledgements{This research has been made possible by the generous support of the National Science Foundation through grants AST 96-19424 and AST 02-04973, and the Wisconsin Space Grant Consortium.}

\end{document}